\documentclass[10pt, conference, compsocconf]{IEEEtran}

\usepackage[T1]{fontenc}
\usepackage[utf8]{inputenc}
\usepackage{authblk}

\usepackage{amsmath}

\usepackage{algorithm}
\usepackage{algorithmic}

\usepackage{comment}

\usepackage{url}

\usepackage[pdftex]{graphicx}
\usepackage{subfigure}
\usepackage{epstopdf}

\begin{document}
%
\title{Variational Inference For Probabilistic Latent Tensor Factorization \\
with KL Divergence}


\author[1]{Beyza Ermi\c{s}\thanks{beyza.ermis@boun.edu.tr}}
\author[1]{Y. Kenan Y{\i}lmaz\thanks{kenan@sibnet.com.tr}}
\author[1]{A. Taylan Cemgil\thanks{taylan.cemgil@boun.edu.tr}}
\author[2]{Evrim Acar\thanks{evrim@life.ku.dk}}
\affil[1]{\textit{Department of Computer Science, Bo\u{g}azi\c ci University, Istanbul, Turkey}}
\affil[2]{\textit{Faculty of Life Sciences, University of Copenhagen, Frederiksberg C, Denmark}}
\renewcommand\Authands{ and }

\maketitle

\begin{abstract}
Probabilistic Latent Tensor Factorization (PLTF) is a recently proposed probabilistic framework for modelling multi-way data. Not only the common tensor factorization models but also any arbitrary tensor factorization structure can be realized by the PLTF framework. This paper presents full Bayesian inference via variational Bayes that facilitates more powerful modelling and allows more sophisticated inference on the PLTF framework. We illustrate our approach on model order selection and link prediction.  

\end{abstract}


\IEEEpeerreviewmaketitle


\section{Introduction}
\label{sec:intro}

Factorization based data modelling has become popular together with the advances in the computational power. Non-negative Matrix Factorization (NMF) model, proposed by Lee and Seung \cite{Lee_Seung_1999} (and also earlier by Paatero and Tapper \cite{PaTa94}), is one of the most popular factorization models where the aim is to estimate the  matrices $Z_1$ and $Z_2$ as the matrix $X$ is observed:
\begin{align}
	X(i,j) \approx \hat{X}(i,j) = \sum_{k} Z_{1}(i,k) Z_{2}(k,j). \label{eqn:NMF}
\end{align}

Here $X$, $Z_1$ and $Z_2$ are all non-negative matrices. This modelling paradigm has found place in many fields including recommender systems \cite{KoBeVo09}, image processing \cite{cemgil09-nmf} and bioinformatics \cite{cichocki09}.

Although the NMF model has its own advantages, certain applications require more structured modelling and incorporation of prior knowledge where NMF can be inadequate. Accordingly, several complex factorization models have been proposed in the literature \cite{cichocki09}. The probabilistic Latent Tensor Factorization framework (PLTF) \cite{Yilmaz:2010} enables one to incorporate domain specific information to any arbitrary factorization model and provides the update rules for multiplicative gradient descent and expectation-maximization algorithms.

The PLTF framework is defined as a natural extension of the matrix factorization model of (\ref{eqn:NMF}):
\begin{align}
	X(v_0)\approx\hat{X}(v_0)=\sum_{\bar{v}_0}\prod_\alpha Z_\alpha(v_\alpha), \label{eqn:tensorFact}
\end{align}
where $\alpha = 1,... K$ denotes the factor index. In this framework, the goal is to compute an approximate factorization of a given higher-order tensor, i.e., a multiway array, $X$ in terms of a product of individual factors $Z_\alpha$, some of which are possibly fixed. Here, we define $V$ as the set of all indices in a model, $V_0$ as the set of visible indices, $V_\alpha$ as the set of indices in $Z_\alpha$, and $\bar{V}_\alpha = V- V_\alpha$ as the set of all indices not in $Z_\alpha$. We use small letters as $v_\alpha$ to refer to a particular setting of indices in $V_\alpha$. Since the product $\prod_\alpha Z_\alpha(v_\alpha)$ is collapsed over a set of indices, the factorization is latent. 

In this study, we use non-negative variants of the two most widely-used low-rank tensor factorization models; the Tucker model \cite{Tu66} and the more restricted CANDECOMP/PARAFAC (CP) model \cite{Ha70, CaCh70, Hi27a}. In order to illustrate the approach, we can define these models in the PLTF notation. Given a three-way tensor $X$ the CP model is defined as follows:
\begin{align}
    X(i,j,k) &\approx\hat{X}(i,j,k)=\sum_r Z_1(i,r)Z_2(j,r)Z_3(k,r) \label{eq:CP}
\end{align}
where the index sets $V=\{i,j,k,r\}$, $V_0=\{i,j,k\}$, $V_1=\{i,r\}$, $V_2=\{j,r\}$ and $V_3=\{k,r\}$.
An alternative Tucker model of $X$ is defined in the PLTF notation as follows:
\begin{align}
    \hat{X}(i,j,k) &=\sum_{p,q,r} Z_1(i,p)Z_2(j,q)Z_3(k,r)Z_4(p,q,r) \label{eq:Tucker2}
\end{align}
where the index sets $V=\{i,j,k,p,q,r\}$, $V_0=\{i,j,k\}$, $V_1=\{i,p\}$, $V_2=\{j,q\}$, $V_3=\{k,r\}$ and $V_4=\{p,q,r\} $.

\vspace{2mm}
The main contributions of this paper can be summarized as follows:
\begin{itemize}
\item Variational Bayes procedure for making inference on the PLTF framework is presented.
\item Exact characterization of the approximating distribution and full conditionals are observed as a product of multinomial distributions, leading to a richer approximation distribution than a naive mean field. 
\item Computation of a variational lower bound for estimation of marginal likelihood of a tensor factorization model is described.
\item A model selection framework for arbitrary non-negative tensor factorization model for KL cost with the variational bound is constructed.
\item The proposed approach is illustrated on link prediction problem: the problem of predicting the existence of connections between entities of interest. 
\end{itemize}

\subsection{Probability Model}

The usual approach to estimate the factors $Z_\alpha$ is trying to find the optimal 
$Z^{*}_{1:K} = {\operatorname{argmin}}_{Z_{1:K}} \hspace{1mm} d(X || \hat{X})$, where $d(.)$ is a divergence typically taken as Euclidean, Kullback-Leibler or Itakura-Saito divergences. Since the analytical solution for this problem is intractable, one should refer to iterative or approximate inference methods. 
\begin{figure}[h!]
\centering
\includegraphics[scale=1.0]{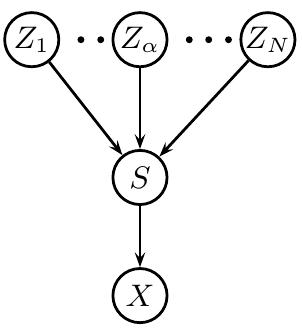}
\caption{The generative model of the PLTF framework as a Bayesian network. The directed acyclic graph describes the dependency structure of the variables: the full joint distribution can be written as $p(X,S,Z_{1:K}) = p(X|S)p(S|Z_{1:K})\prod_\alpha p(Z_\alpha)$}.
\label{fig:bayesianNetwork}
\end{figure}

In this study, we use the Kullback-Leibler (KL) divergence as the cost function which is equivalent to selecting the Poisson observation model \cite{cemgil09-nmf,Yilmaz:2010}, while our approach can be extended to other costs where a composite structure is present. The overall probabilistic model is defined as follows:
\begin{align}
  \tag{factor priors}
  Z_\alpha(v_\alpha) &\sim \mathcal{G}(Z_\alpha(v_\alpha); A_\alpha(v_\alpha), B_\alpha(v_\alpha))   \label{eqn:factorPriors} \\
  \tag{intensity}  
  \Lambda(v) &= \prod_\alpha Z_\alpha(v_\alpha) \label{eqn:intensity} \\
  \tag{KL-cost}
  S(v) &\sim \mathcal{PO}(S(v); \Lambda(v)) \label{eqn:KL-cost}  \\
  \tag{observation}
  X(v_0) &= \sum_{\bar{v}_0} S(v) \label{eqn:observation} \\
  \tag{parameter}
  \hat{X}(v_0) &= \sum_{\bar{v}_0} \Lambda(v) \label{eqn:parameter} 
\end{align}
where the symbols refer to Poisson and Gamma distributions respectively, where:
\begin{align}
	\mathcal{PO}(s;\lambda) &= e^{-\lambda} \frac{\lambda^{s}}{s!}	\label{eqn:poisson_dist} \\
	\mathcal{G}(z;a,b) &= e^{-bz}	\frac{z^{a-1}b^{a}}{\Gamma(a)}. \label{eqn:gamma_dist}	 
\end{align}

The Gamma prior on the factors are chosen in order to preserve conjugacy. The graphical model for the PLTF framework is depicted in Figure~\ref{fig:bayesianNetwork}. Note that $p(X|S)$ is a degenerate distribution that is defined as follows:
\begin{align}
	p(X|S) =  \prod_{v_0} \delta \left(X(v_0) - \sum_{\bar{v}_0} S(v) \right). \label{eqn:deg_dist}
\end{align}

Here, $\delta(.)$ is the Kronecker delta function where $\delta(x) = 1$ when $x = 0$ and $\delta(x) = 0$ otherwise.

\vspace{2mm}
\subsubsection*{Missing data}
To model missing data, we define a $0-1$ mask array $M$, the same size as $X$ where $M(v_{0})=1$ ($M(v_{0})=0$) if $X(v_{0})$ is observed (missing). Using the mask variables, the missing data is handled smoothly by the following observation model in PLTF:
\begin{align}
\scalebox{0.92}{$p(X|S)p(S|Z_{1:N}) = \prod_{v_0}\prod_{\bar{v}_0}\lbrace p(X(v_0)|S(v))p(S(v)|Z_{1:N})\rbrace^{M(v_0)}$}
\label{eqn:handle_missing}
\end{align} 
where slight modifications are needed to be done in VB based update equation is shown in section\ref{sec:missing_data}.

\subsection{Fixed Point Update Equation for $PLTF_{KL}$}

Here, we recall the generative Probabilistic Latent Tensor Factorization KL model $(PLTF_{KL})$\ref{eqn:factorPriors} with the following fixed point iterative update equation for the component $Z_{\alpha}$ obtained via EM as:
\begin{align}
\scalebox{1.02}{$Z_{\alpha}(v_{\alpha}) \leftarrow \frac{(A_{\alpha}(v_{\alpha})-1)+Z_{\alpha}(v_{\alpha})\sum_{\bar{v}_{\alpha}}M(v_0)\frac{X(v_0)}{\hat{X}(v_0)}\prod_{\alpha\prime\neq\alpha}Z_{\alpha\prime}(v_{\alpha\prime})}{\frac{A_{\alpha}(v_{\alpha})}{B_{\alpha}(v_{\alpha})}+\sum_{\bar{v}_{\alpha}}M(v_0)\prod_{\alpha\prime\neq\alpha}Z_{\alpha\prime}(v_{\alpha\prime})}$}
\label{eqn:updateEq_EM}
\end{align} 
where $\hat{X}(v_0)$ is the model estimate defined as earlier $\hat{X}(v_0)=\sum_{\bar{v}_{0}}\prod_{\alpha}Z_{\alpha}(v_{\alpha})$. We note that the gamma hyperparameters $A_\alpha(v_\alpha)$ and $B_\alpha(v_\alpha)/A_\alpha(v_\alpha)$ are chosen for computational convenience for sparseness representation such that the distribution has a mean $B_{\alpha}(v_{\alpha})$ and standard deviation $B_{\alpha}(v_{\alpha})/\sqrt{A_{\alpha}(v_{\alpha})}$ and for small $A_{\alpha}(v_{\alpha})$ most of the parameters are forced to be around $0$ favoring for a sparse representation \cite{cemgil09-nmf}. So, equation(\ref{eqn:updateEq_EM}) can be approximated as:
\begin{align}
	Z_{\alpha}(v_{\alpha}) \leftarrow \frac{\sum_{\bar{v}_{\alpha}}M(v_0)\frac{X(v_0)}{\hat{X}(v_0)}\prod_{\alpha\prime\neq\alpha}Z_{\alpha\prime}(v_{\alpha\prime})}{\sum_{\bar{v}_{\alpha}}M(v_0)\prod_{\alpha\prime\neq\alpha}Z_{\alpha\prime}(v_{\alpha\prime})}
\label{eqn:updateEq_EM2}
\end{align} 

\vspace{2mm}
\subsubsection*{Tensor forms via $\Delta$ function}
We make use of $\Delta$ function to make the notation shorter and implementation friendly. A tensor valued $\Delta_{\alpha}^{Z}(Q)$ function associated with component $Z_\alpha$ is defined as follows:
\begin{align}
   \Delta_{\alpha}^{Z}(Q) = \left[ \sum_{\bar{v}_\alpha} \left( Q(v_0) \prod_{\alpha\prime\neq\alpha} Z_\alpha\prime(v_\alpha\prime)\right) \right]    \label{eqn:delta_func} 
\end{align}
Recall that $\Delta_{\alpha}^{Z}(Q)$ is an object the same size of $Z_\alpha$ while $\Delta_{\alpha}^{Z}(Q)(v_\alpha)$ refers to a particular element of $\Delta_{\alpha}^{Z}(Q)$.

Now, equation(\ref{eqn:updateEq_EM2}) can be written into a form that by use of $\Delta_{\alpha}^{Z}(.)$ as: 
\begin{align}
	Z_\alpha \leftarrow Z_\alpha \circ \Delta_{\alpha} (M \circ X/\hat{X})/\Delta_{\alpha}(M)  \label{eqn:updateEq_EM3}
\end{align}
where as usual $\circ$ and $/$ stand for element wise multiplication(Hadamard product) and division respectively. We use update equation (\ref{eqn:updateEq_EM3}) in the following chapters for PLTF-EM method to compare with the PLTF-VB method.

\section{Variational Bayes}
\label{sec:method}

For a Bayesian point of view, a model is associated with a random variable $\Theta$ and interacts with the observed data X simply as $p(\Theta|X) \propto p(X|\Theta)p(\Theta)$. The quantity $p(X|\Theta)$ is called \textit{marginal likelihood} \cite{bishop06} and it is average over the space of the parameters, in our case, $S$ and $Z$ as \cite{cemgil09-nmf}.
\begin{align}
	p(X|\Theta) = \int_Z dZ \sum_S p(X|S,Z,\Theta) p(S,Z|\Theta) \label{eqn:marginal_lh}
\end{align}

On the other hand, computation of this integral is itself a difficult task that requires averaging on several models and parameters. There are several approximation methods such as sampling or deterministic approximations such as Gaussian approximation. One other approximation method is to bound the log marginal likelihood by using \textit{variational inference} \cite{bishop06,cemgil09-nmf,GhahramaniB00} where an approximating distribution $q$ is introduced into the log marginal likelihood equation:
\begin{align}
	\log p(X|\Theta) \geq \int_Z dZ \sum_S q(S|Z) \log \frac{p(X,S,Z|\Theta)}{q(S,Z)}   \label{eqn:app_dist}
\end{align}
where the bound attains its maximum and becomes equal to the log marginal likelihood whenever $q(S,Z)$ is set as $p(S,Z|X,\Theta)$, that is the exact posterior distribution. However, the posterior is usually intractable, and rather, inducing the approximating distribution becomes easier. Here, the approximating distribution $q$ is chosen such that it assumes no coupling between the hidden variables such that it factorizes into independent distributions as $q(S,Z) = q(S) q(Z)$. As exact computation is intractable, we will resort to standard variational Bayes approximations \cite{bishop06,GhahramaniB00}. The interesting result is that we get a belief propagation algorithm for marginal intensity fields rather than marginal probabilities.

\subsection{Variational Update Equations for $PLTF_{KL}$}
\label{sec:var_update_eq}

Here, we formulate the fixed point update equation for the update of the factor $Z_{\alpha}$ as an expectation of the approximated posterior distribution \cite{YKYthesis}. Approximation for posterior distribution $q(Z)$ is identified as the gamma distribution with the following parameters:
\begin{align}
	Z_\alpha(v_\alpha) \sim \mathcal{G}(Z_\alpha(v_\alpha); C_\alpha(v_\alpha), D_\alpha(v_\alpha)) \label{eqn:appPosterior} 
\end{align} 
where the shape and scale parameters are:
\begin{align}
	C_\alpha(v_\alpha) &= A_\alpha(v_\alpha) + \sum_{\bar{v}_\alpha}\frac{X_(v_0)}{\hat{X}_L(v_0)}\prod_\alpha L_\alpha(v_\alpha)  \label{eqn:hyperC}  \\
	D_\alpha(v_\alpha) &= \left( \frac{A_{\alpha}(v_\alpha)}{B_{\alpha}(v_\alpha)}  + \sum_{\bar{v}_\alpha} \prod_{\alpha\prime\neq\alpha} \langle Z_{\alpha\prime}(v_{\alpha\prime}) \rangle \right)^{-1}   \label{eqn:hyperD}  	
\end{align}

Hence the expectation of the factor $Z_{\alpha}$ is identified as the mean of the gamma distribution and given in the iterative fixed point update equation obtained via variational Bayes:
\begin{align}
\langle Z_{\alpha}(v_{\alpha}) \rangle &= C_\alpha(v_\alpha) D_\alpha(v_\alpha) \label{eqn:IFPUE}  \\   
&= \frac{A_{\alpha}(v_{\alpha})+L_{\alpha}(v_{\alpha})\sum_{\bar{v}_{\alpha}}\frac{X(v_0)}{\hat{X}_L(v_0)} \prod_{\alpha\prime\neq\alpha} L_{\alpha\prime}(v_{\alpha\prime})}{\frac{A_{\alpha}(v_{\alpha})}{B_{\alpha}(v_{\alpha})} + \sum_{\bar{v}_{\alpha}}\prod_{\alpha\prime\neq\alpha} E_{\alpha\prime}(v_{\alpha\prime})} \label{eqn:updateEq_VB}
\end{align} 
$E_{\alpha}(v_{\alpha})$ and $L_{\alpha}(v_{\alpha})$ ($L$ due to `Log') are two forms of expectations of $Z_{\alpha}(v_{\alpha})$ while $\hat{X}_E(v_0)$ and $\hat{X}_L(v_0)$ are model outputs generated by the components $E_{\alpha}(v_{\alpha})$ and $L_{\alpha}(v_{\alpha})$. While $\hat{X}_E$ is not being used in Equation(\ref{eqn:updateEq_VB}) we define it here, in addition to $\hat{X}_L$, (and use it later on) since $\hat{X}_E$ has the same shape as $\hat{X}_L$. Indeed $\hat{X}_E$ and $\hat{X}_L$ can be regarded as different `views' of $\hat{X}$ since they have the same shape (dimensions) as $\hat{X}$ and their computations are done via the same matrix primitives as $\hat{X}$. Here:
\begin{align}
	E_{\alpha}(v_{\alpha}) &= \langle Z_{\alpha}(v_{\alpha}) \rangle = C_{\alpha}(v_{\alpha}) D_{\alpha}(v_{\alpha})   \label{eqn:expE}  \\
	L_{\alpha}(v_{\alpha}) &= \exp\left(\langle \log Z_{\alpha}(v_{\alpha}) \rangle \right) = \exp \left( \psi \left(C_{\alpha}(v_{\alpha}) \right) \right) D_{\alpha}(v_{\alpha})   \label{eqn:expL}  \\
	\hat{X}_E(v_0) &= \sum_{\bar{v}_0} \prod_{\alpha} E_{\alpha}(v_{\alpha})   \label{eqn:XexpE}  \\
	\hat{X}_L(v_0) &= \sum_{\bar{v}_0} \prod_{\alpha} L_{\alpha}(v_{\alpha})   \label{eqn:XexpL}  
\end{align} 

Note that the $VB$ version of the update equation(\ref{eqn:updateEq_VB}) closely resembles the $EM$ version given in (\ref{eqn:updateEq_EM}). Indeed when the observed values are large, digamma function becomes $\lim_{x\rightarrow\infty}\psi(x)/\log(x)=1$, and this, in turn, gives $L_\alpha(v_\alpha) \simeq E_\alpha(v_\alpha)$ and $\hat{X}_L(v_0) \simeq \hat{X}_E(v_0)$.

\subsection{Variational Bound and Sufficient Statistics}

The marginal likelihood of the observed data under a tensor factorization model $p(X)$ is often necessary for certain problems such as model selection. We lower bound the marginal likelihood for any arbitrary $PLTF_{KL}$ model based on variational Bayes; while clearly other Bayesian model selection such as MCMC \cite{cemgil09-nmf} can also be used. To bound the marginal log-likelihood, an approximating distribution $q(S,Z)$ over the hidden structure $S$ and $Z$ is introduced as:
\begin{align}
	\mathcal{L}(\Theta) &= \log p(X|\Theta) \geq \int_Z dZ \sum_S q(S,Z) \log \frac{p(X,S,Z|\Theta)}{q(S,Z)}   \label{eqn:bound}  \\
	&= \langle \log p(X,S,Z|\Theta) \rangle_{q(S,Z)} + H\left[ q(S,Z)\right] = \mathcal{B}_{VB}[q]  \label{eqn:bound2} 
\end{align}

The bound is tight whenever $q$ equals to the posterior as $q(S,Z)=p(S,Z|X,\Theta)$ but computing the posterior $p(S,Z|X,\Theta)$ is intractable. At this point variational Bayes suggests approximating $q$. The simplest selection for $q$ from the family of approximating distribution is the one which poses no coupling for the members of the hidden structure $S$, $Z$. That is, we take a factorized approximation $q(S,Z)=q(S)q(Z) $ such that:
\begin{align}
	q(S,Z) = \left( \prod_{v_0} q\left(S(v_0,*)\right) \right) \left( \prod_{\alpha} \prod_{v_0} q\left(Z_{\alpha}(v_{\alpha})\right) \right)   \label{eqn:factorized_app} 
\end{align}
where $*$ symbol in $S(v_0,*)$ is used to indicate the slice of the array. That is $S(v_0,*)$ is the slice of the latent tensor $S$ as the observed variables in configurations $v_0$ are being fixed. Then, we have:

\begin{align}
	q_{S(v_0,*)}^{(n+1)} &\propto \exp\left(\langle \log p(X,S,Z|\Theta)\rangle_{q^{(n)}/q_{S(v_0,*)}} \right)  \label{eqn:qs} \\
	q_{Z_\alpha(v_\alpha)}^{(n+1)} &\propto \exp\left(\langle\log p(X,S,Z|\Theta)\rangle_{q^{(n+1)}/q_{Z_{\alpha}(v_\alpha)}}\right)   
	\label{eqn:qz}	 
\end{align}
where the superscript $(n)$ indicates the iteration index. This iteration monotonically improves the individual factors of the $q$ distribution, that is, $\mathcal{B}[q^{(n)}]\leq\mathcal{B}[q^{(n+1)}]$ for $n=1,2,...$ given an initialization $q^{(0)}$. 

First, we start with formulating the approximating distribution $q(S)$. When we expand the $\log$ and drop $\log P(Z|\Theta)$ and all other irrelevant $S$ terms $q_{S(v_0,*)}$ we end up with:
\begin{align}
	q_{S(v_0,*)} \propto \exp\left(\langle\log p(X|S)+\log p(S|Z)\rangle_{q/q_{S(v_0,*)}} \right)  \label{eqn:qs_expanded} \\
	\propto \exp\left(\sum_{\bar{v}_0} \left( S(v)\langle \log\prod_{\alpha} Z_{\alpha}(v_\alpha)\rangle - \log\Gamma(S(v)+1)\right)  \right.  \notag\\  
	\left. + \log \delta \left( X(v_0) - \sum_{\bar{v}_0} S(v)\right)\right) \label{eqn:qs_expanded2}   \\
	\propto \exp\left(\sum_{\bar{v}_0} \left( S(v) \sum_{\alpha} \log \langle Z_{\alpha}(v_\alpha)\rangle - \log\Gamma(S(v)+1)\right)\right)   \notag\\  
	+ \delta \left( X(v_0) - \sum_{\bar{v}_0} S(v) \right) \label{eqn:qs_expanded3} 	                  
\end{align}

Exactly, the slice $S(v_0,*)$ is sampled from the multinomial distribution as $X(v_0)$ is the total number of observations. Here, $s$ is a vector of a priori independent Poisson random variables $s_i$. $\lambda$ is intensity vector conditioned on the sum $x=\Sigma_i s_i$ and $x$ is multinomial distributed with cell probabilities $p=\lambda/\Sigma_i\lambda_i$. The joint posterior density of $s$ is denoted by $\mathcal{M}(s;x,p)$. Finally we obtain the approximating distribution as:  
\begin{align}
	q_{S(v_0,*)} \sim \mathcal{M}(S(v_0,*), X(v_0), P(v_0,*))  \label{eqn:qs_final} 
\end{align}
Then, the cell probabilities and sufficient statistics for $q_{S(v_0,*)}$ are:
\begin{align}
	P(v) &= \frac{\exp(\Sigma_\alpha\langle\log Z_\alpha(v_\alpha)\rangle)}{\Sigma_{\bar{v}_0}\exp(\Sigma_\alpha\langle\log Z_\alpha(v_\alpha)\rangle)}   \label{eqn:cell_prob}  \\
	\langle S(v)\rangle &= X(v_0)P(v)  \label{eqn:sufficient_stat} 
\end{align}

Now, we turn to formulating $q(Z)$. The distribution $q_{Z_\alpha(v_\alpha)}$ is obtained similarly. After we expand the $\log$ and drop irrelevant terms, it becomes proportional to:
\begin{align}
q_{Z_{\alpha}(v_\alpha)} &\propto \exp \left( \langle \log p(S|Z) + \log p(Z|\Theta)\rangle_{q/q_{Z_{\alpha}(v_\alpha)}} \right) \label{eqn:qz2} \\
&\propto \log Z_\alpha(v_\alpha) \left( A_\alpha(v_\alpha)-1 + \sum_{\bar{v}_\alpha}\langle S(v)\rangle \right)    \notag\\  
& -Z_\alpha(v_\alpha) \left( \frac{A_\alpha(v_\alpha)}{B_\alpha(v_\alpha} + \sum_{\bar{v}_\alpha} \prod_{\alpha\prime\neq\alpha} \langle Z_{\alpha\prime}(v_{\alpha\prime})\rangle \right) \label{eqn:qz3}
\end{align}
which is the distribution
\begin{align}
	q_{Z_{\alpha}(v_\alpha)}  \sim \mathcal{G}(C_\alpha(v_\alpha), D_\alpha(v_\alpha))   \label{eqn:qfinal}
\end{align}
where the shape and scale parameters for $q_{{Z_\alpha(v_\alpha)}}$ are given in equation~(\ref{eqn:hyperC}) and equation~(\ref{eqn:hyperD}).

Finally, sufficient statistics are obtained by the definition of the gamma distribution as follows:
\begin{align}
	E_\alpha(v_\alpha) &= \langle Z_\alpha(v_\alpha)\rangle =  C_\alpha(v_\alpha) D_\alpha(v_\alpha)  \label{eqn:resE}  \\
	L_\alpha(v_\alpha) &= \exp(\langle\log Z_\alpha(v_\alpha) \rangle) = \exp(\psi(C_\alpha(v_\alpha)))D_\alpha(v_\alpha)  \label{eqn:resL}	
\end{align}

\subsubsection*{Handling Missing Data}
\label{sec:missing_data}

Here, slight modifications are needed in the VB-based update equation. We start with the modification on the full joint. Priors are not part of the observation model so they are not affected. The first two terms of $\langle \log p(X,S,Z|\Theta)\rangle_{q(S,Z)}$ become:
\begin{align}
	& \sum_{v_0}M(v_0)\left\langle \log \delta \left( X(v_0)-\sum_{\bar{v}_0}S(v) \right) \right\rangle     \notag\\
    & + \sum_{v_0}M(v_0) \left( \langle S(v)\rangle \left\langle \log \prod_\alpha Z_\alpha(v_\alpha) \right\rangle  \right.  \notag\\
    & \left. - \prod_\alpha \langle Z_\alpha(v_\alpha)\rangle - \langle \log \Gamma (S(v)+1) \rangle \right) ...  \label{eqn:VB_mask}	
\end{align}
 and this results in the following: 
\begin{align}
	q_{Z_\alpha}(v_\alpha) &\propto \log\langle Z_\alpha(v_\alpha)\rangle\ \left(A_\alpha(v_\alpha)-1 + \sum_{\bar{v}_\alpha}M(v_0)\langle S(v)\rangle \right)   \notag\\
	&-\langle Z_\alpha(v_\alpha)\rangle(\frac{A_\alpha(v_\alpha)}{B_\alpha(v_\alpha)}+\sum_{\bar{v}_\alpha}M(v_0)\prod_{\alpha\prime\neq\alpha}\langle Z_{\alpha\prime}(v_{\alpha\prime})\rangle)  \label{eqn:qz_mask}  \\
	&\propto \mathcal{G}(C_\alpha(v_\alpha), D_\alpha(v_\alpha))   \label{eqn:gamma_mask}	
\end{align}
 
This modifies the gamma parameters for $q(Z)$ given in equations (\ref{eqn:hyperC}) and (\ref{eqn:hyperD}) to include the mask $M(v_0)$ as follows:
\begin{align}
	C_\alpha(v_\alpha) &= A_\alpha(v_\alpha) + \sum_{\bar{v}_\alpha}M(v_0)\langle S(v)\rangle   \label{eqn:hyperC_mask}  \\
	D_\alpha(v_\alpha) &= \left( \frac{A_{\alpha}(v_\alpha)}{B_{\alpha}(v_\alpha)}  + \sum_{\bar{v}_\alpha} M(v_0) \prod_{\alpha\prime\neq\alpha} \langle Z_{\alpha\prime}(v_{\alpha\prime}) \rangle \right)^{-1}   \label{eqn:hyperD_mask} 
\end{align}
 
The other terms are not affected since mask matrix is already in the definition of $C_\alpha(v_\alpha)$ and $D_\alpha(v_\alpha)$. $\hat{X}_E$ and $\hat{X}_L$ are already defined in terms of $C_\alpha(v_\alpha)$ and $D_\alpha(v_\alpha)$. Moreover, $A_\alpha(v_\alpha)$ and $B_\alpha(v_\alpha)$ are priors and not part of the observation model. 


Now, for $C_\alpha(v_\alpha)$, we need to find out $\Sigma_{\bar{v}_\alpha}\langle S(v)\rangle$, which can be written as:
\begin{align}
	\sum_{\bar{v}_\alpha}\langle S(v)\rangle &= \sum_{\bar{v}_\alpha} X(v_0)p(v) = \sum_{\bar{v}_\alpha}\frac{X(v_0)}{\hat{X}_L(v_0)}\prod_{\alpha} L_{\alpha}(v_\alpha)  \\
	&= L_\alpha(v_\alpha) \sum_{\bar{v}_\alpha} \frac{X(v_0)}{\hat{X}_L(v_0)} \prod_{\alpha\prime \neq \alpha} L_{\alpha\prime}(v_{\alpha\prime})     \label{eqn:finalSS} 
\end{align}

After consideration of the missing data for our approach, $C_\alpha$ and $D_\alpha$ can be written using the $\Delta_{\alpha}^{E}(.)$ and $\Delta_{\alpha}^{L}(.)$ as:
\begin{align}
	C_\alpha &= A_\alpha + L_\alpha \circ \Delta_{\alpha}^{L}(M \circ X /\hat{X}_L)    \label{eqn:hyperC_final}  \\
	D_\alpha &= \left(\frac{A_\alpha}{B_\alpha} + \Delta_{\alpha}^{E}(M) \right)^{-1}	 \label{eqn:hyperD_final} 
\end{align}
that, in turn, since $\langle Z_\alpha\rangle$ is $C_\alpha \circ D_\alpha$, $E_\alpha$ and $L_\alpha$ the sufficient statistics for $q(Z_\alpha)$ become:
\begin{align}
	\langle Z_\alpha\rangle &= E_\alpha \leftarrow \frac{A_\alpha + L_\alpha \circ \Delta_{\alpha}^{L}(M \circ X /\hat{X}_L)}{\frac{A_\alpha}{B_\alpha} + \Delta_{\alpha}^{E}(M)}     \label{eqn:Z_final}  \\
	\exp\langle\log(Z_\alpha)\rangle &= L_\alpha \leftarrow \exp(\psi(C_\alpha)) \circ D_\alpha   \label{eqn:expZ_final} 
\end{align}

After straightforward substitutions, we obtain the variational probabilistic latent tensor factorization algorithm, that can compactly be expressed as in Algorithm\ref{alg:pltf_vb}.
 
\begin{algorithm}
\caption{Variational Inference for PLTF (PLTF-VB)}
\label{alg:pltf_vb}
\begin{algorithmic} 
\REQUIRE X (observation), M (mask array), A and B (priors)
\ENSURE E (expected value of factors), and $\mathcal{B}$ (bound)
\STATE Here $N = |\alpha|$
\FOR{$\alpha = 1 ... N$} 
\STATE $L_\alpha \sim \mathcal{G}(A_\alpha, B_\alpha/A_\alpha) $ 
\STATE $E_\alpha \sim \mathcal{G}(A_\alpha, B_\alpha/A_\alpha) $ 
\ENDFOR
\STATE Main loop
\FOR{$epoch = 1 ... MAXITER$} 
\STATE Compute $\hat{X}_L$ and $\hat{X}_E$
\STATE $\hat{X}_L(v_0) = \sum_{\bar{v}_0}\prod_\alpha L_\alpha(v_\alpha)$
\STATE Computation for $\hat{X}_E$ is similar and is omitted
\FOR{$\alpha = 1 ... N$} 
\STATE $C_\alpha = A_\alpha + L_\alpha \circ \Delta_{\alpha}^{L}(M \circ X / \hat{X}_L) $  
\STATE $D_\alpha = 1/((A_\alpha/B_\alpha) + \Delta_{\alpha}^{E}(M)) $
\STATE $E_\alpha = C_\alpha \circ D_\alpha $
\ENDFOR 
\FOR{$\alpha = 1 ... N$} 
\STATE $L_\alpha = \exp(\psi(C_\alpha)) \circ D_\alpha $
\ENDFOR           
\ENDFOR
\end{algorithmic}
\end{algorithm}

\section{Experiments and Results}
\label{sec:experiments}

In this section, we demonstrate the use of the proposed variational Bayesian PLTF (PLTF-VB) for model selection and missing link prediction. First, we study model selection on synthetic datasets and show that the proposed approach can accurately determine the number of components in a CP model. We also show the performance of PLTF-VB for model selection on a real data set, i.e., the UCLAF \cite{ZhCaZhXiYa10}. Furthermore, on the UCLAF dataset, we study the missing link prediction problem and compare the performance of the proposed variational Bayesian PLTF (PLTF-VB) with the standard PLTF (PLTF-EM) in terms of missing link prediction recovery. For the experiments we use the algorithm that implements variational fixed point update equation given in panel Algorithm~\ref{alg:pltf_vb} and we use the equation given in (\ref{eqn:bound2}) for variational bound computation. 

\vspace{2mm}
\subsubsection*{Data}
As the real data, we use the UCLAF dataset \footnote{\url{http://www.cse.ust.hk/~vincentz/aaai10.uclaf.data.mat}} \cite{ZhCaZhXiYa10} extracted from the GPS data that include information of three types of entities: user, location and activity. The relations between the user-location-activity triplets are used to construct a three-way tensor $X$. In tensor $X$, an entry $X(i,j,k)$ indicates the frequency of a user $i$ visiting location $j$ and doing activity $k$ there; otherwise, it is $0$. Since we address the link prediction problem in this study, we define the user-location-activity tensor $X$ as:
 \begin{align}
  X(i,j,k) = 
   \begin{cases}
      1 & \text{if user $i$ visits location $j$ and} \\
       & \text{ performs activity $k$ there,} \\
      0 & \text{otherwise.}
   \end{cases} 
\end{align}

To construct the dataset, the raw GPS points were clustered into 168 meaningful locations and the user comments attached to the GPS data were manually parsed into activity annotations for the 168 locations. Consequently, this dataset consists of 164 users, 168 locations and 5 different types of activities, including `Food and Drink', `Shopping', `Movies and Shows', `Sports and Exercise', and `Tourism and Amusement'. In this dataset, 18 users have no location and activity information; it means that the slices corresponding to these users are completely missing. Therefore, we have only used the data from the remaining 146 users. So in our experiments, the number of users is $I=146$, the number of locations $J=168$ and the number of activities $K=5$.

\vspace{2mm}
\subsubsection*{Computational Environment}

All experiments were performed using MATLAB 2010b on 2.4GHz Core i5 520M processor and 4GB RAM. Timings were performed using MATLAB's tic and toc functions.

\subsection{Model Selection}

Using both synthetic datasets and the UCLAF dataset, we assess the performance of our approach in a model selection context where the goal is to determine the cardinality of the latent index $r$ of the CP model $X^{i,j,k} = \sum_r A^{i,r}B^{j,r}C^{k,r}$ . We denote the cardinality of an index $i$ as $|i|$. $|r|$ is set to be from $2$ to $10$ (ignoring $1$) incremented by $1$ gradually at each run. In the experiments, the iteration number is set to $2000$, the shape parameter $A$ and the scale parameter $B$ of the gamma priors are set to be $0.5$ and $10$ respectively. As an initialization, a number of random initializations, i.e., $10$, are used and the best performing one is picked.


Third-order tensors of different sizes following a CP model are generated. The cardinality of the observed indices $i \times j \times k$ (i.e., the size of the data) are set to $50 \times 50 \times 50$ and $500 \times 500 \times 500$. The cardinality of the latent index $r$ is set to $7$ as the true model order. Each run is repeated $10$ times and average bound score is plotted in the figures as model order is on the x-axis while bound score given in Equation(\ref{eqn:bound}) on the y-axis.  
   
\begin{figure}[h!]
\begin{minipage}[b]{0.5\textwidth}
\centering
  \subfigure[Missing data case with synthetic data]{\label{subfig:model_determination_synth}\includegraphics[scale=0.45]{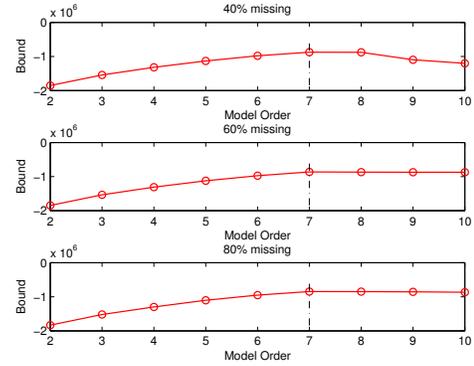}} \\
  \subfigure[Missing data case with UCLAF]{\label{subfig:model_determination_missing}\includegraphics[scale=0.45]{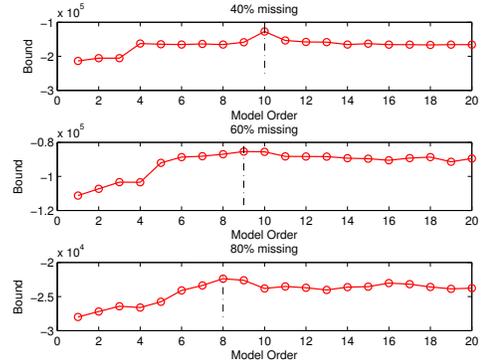}}
  \caption{Model order selection using variational bound for CP generated data}
  \label{fig:model_selection}
\end{minipage}
\end{figure}

In the first experiment we use the dataset with the size of $50 \times 50 \times 50$ to test model selection process when $40\%$, $60\%$ and $80\%$ of data is unobserved respectively. What we expect to see on Figure~\ref{subfig:model_determination_synth} is simply that at around true model order of $|r|=7$ the bound to be the highest to demonstrate that PLTF-VB can find the true model order correctly in all cases of the presence of missing data.

In the second experiment, to illustrate the model order selection performance under missing data case of our model with real data, we use UCLAF dataset. As for the experiment settings, the gamma hyperparameters are set to $0.5$ for scale and $10$ for shape for all the components and $|r|$ is set to be from $1$ to $20$. Figure~\ref{subfig:model_determination_missing} shows the performance of our model with $40\%$, $60\%$ and $80\%$ missing data. We can see from this figure that when the amount of missing data increases the best model order decreases.


\subsection{Scalability}

\begin{figure}[h!]
\begin{minipage}[b]{0.5\textwidth}
\centering
\includegraphics[scale=0.45]{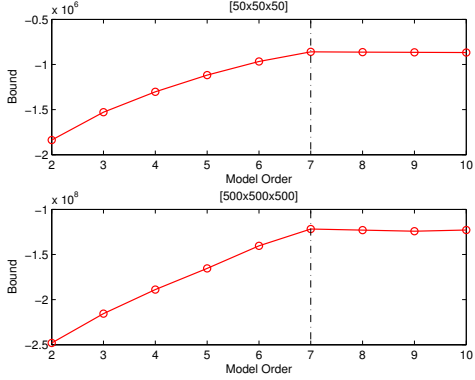}
\caption{Model order selection on datasets with different sizes}
\label{fig:model_determination_scalable}
\end{minipage}
\end{figure}

To test the scalability of the PLTF-VB approach, we have done experiments by using the datasets with the size of $50 \times 50 \times 50$ and $500 \times 500 \times 500$. We evaluate the results in terms of both accuracy and speed.  Figure~\ref{fig:model_determination_scalable} shows that PLTF-VB algorithm determines the true model order in both datasets correctly.
Then, we run the PLTF-VB algorithm on both datasets ten times and obtain the average solve times. In the $500 \times 500 \times 500$  case, the solve time takes 3974 seconds, approximately 1000 times slower than the $50 \times 50 \times 50$ case (its solve time takes 37 seconds in average), which had 1/1000 times as many variables.
In each iteration, the complexity of Algorithm~\ref{alg:pltf_vb} is $O(IJKR)$ where $I$,$J$,$K$ are the cardinality of the observed indices and $R$ is the cardinality of the latent index. Consequently, this experiment demonstrates that size of the data and algorithm's complexity are linearly correlated, when one increases the other also increases by the same amount. 

\subsection{Hyperparameter Selection}

\begin{figure}[h!]
\begin{minipage}[b]{0.5\textwidth}
\centering
	\subfigure[40\% missing]{
	\includegraphics[scale=0.45]{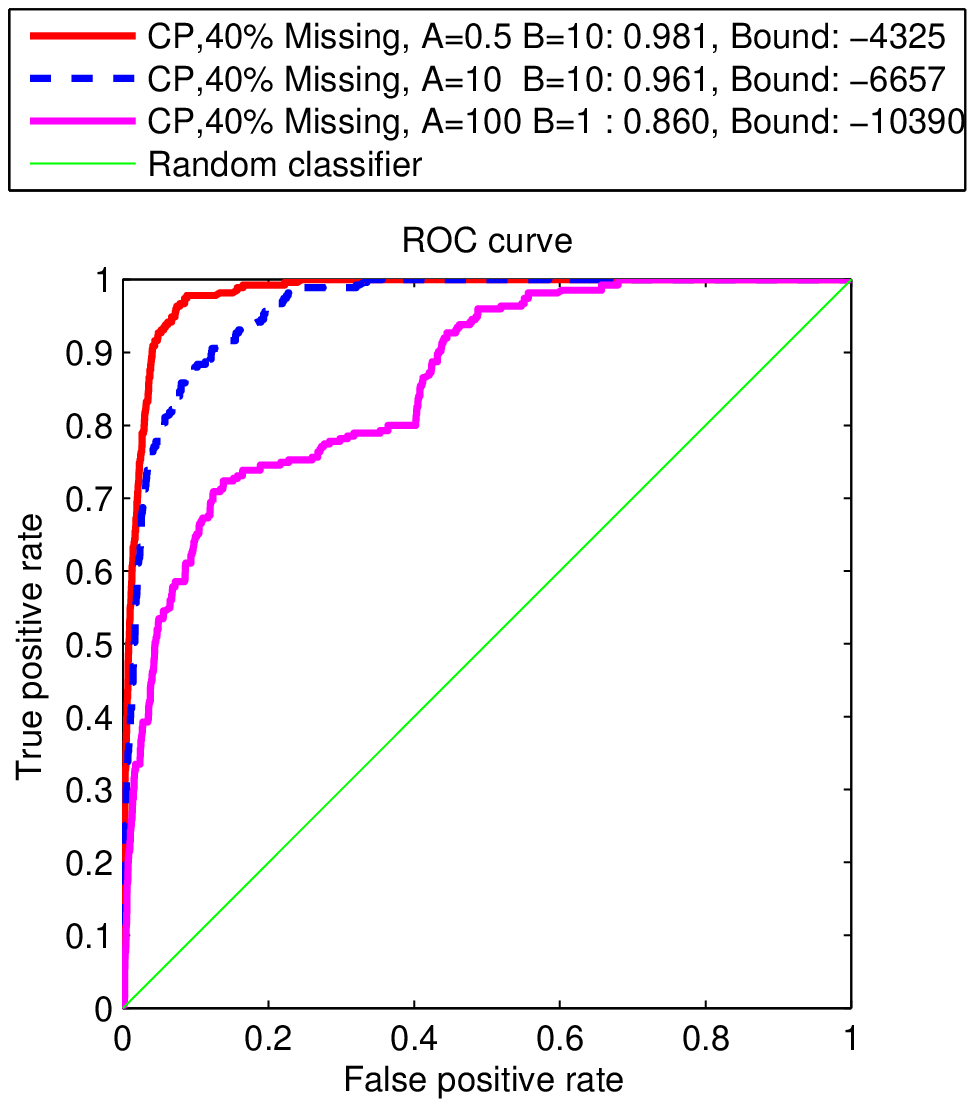}\label{fig:subfig1}}\\
	\subfigure[80\% missing]{
	\includegraphics[scale=0.45]{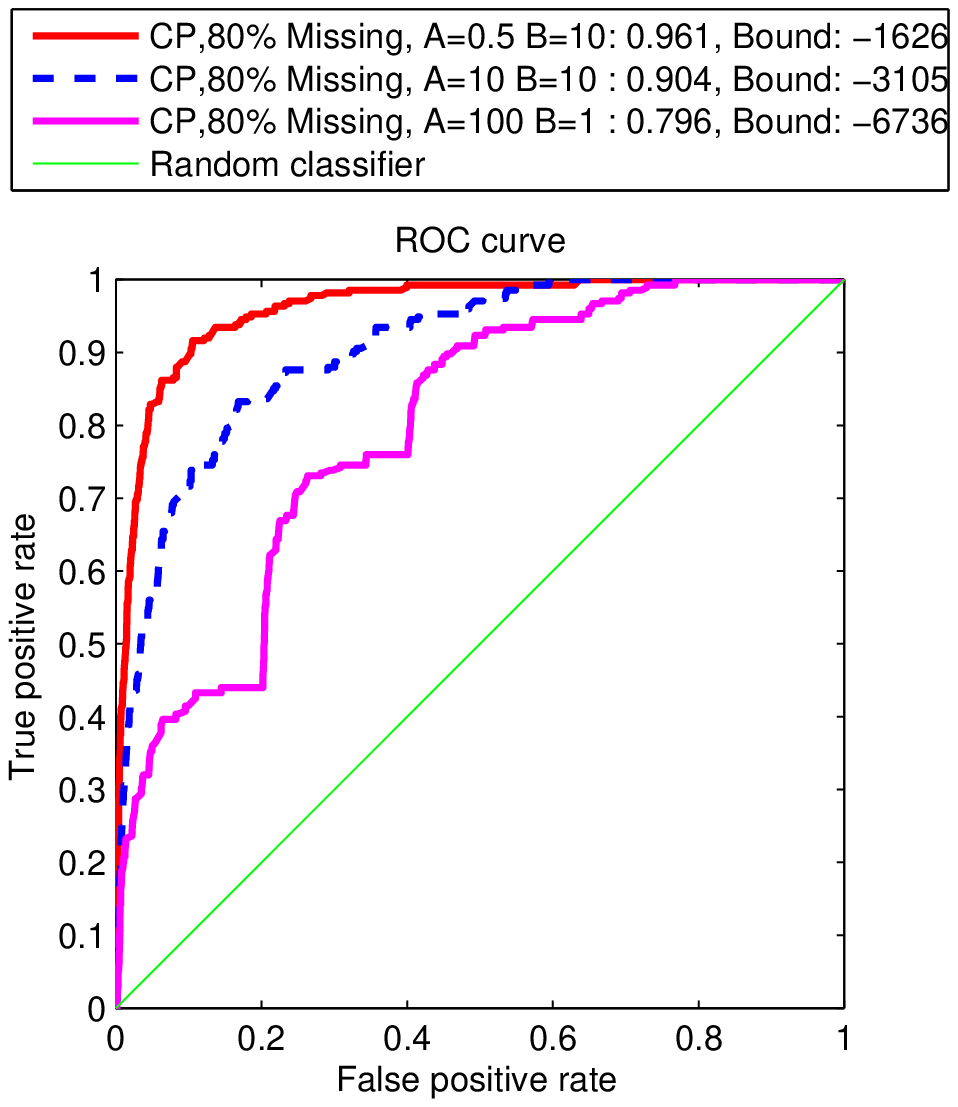}\label{fig:subfig2}}
\caption{Effect of hyperparameter selection with CP model when R=2.}
\label{fig:compHyperparameters}
\end{minipage}
\end{figure}

We observe that hyperparameter adaptation is crucial for obtaining good prediction performance. In our simulations, results for PLTF-VB without hyperparameter adaptation were occasionally poorer than the PLTF-EM estimates. We set both shape $A$ and scale $B$ hyperparameters same for all components $Z_{1:3}$. We tried several number of different values for hyperparameters to obtain the best prediction results under missing data case. Figure~\ref{fig:compHyperparameters} shows the comparison of three different hyperparameter settings; $A=0.5,B=10$, $A=10,B=10$ and $A=100,B=1$ in terms of link prediction performance. As we can see, we obtain best result when initialising the shape hyperparameter $A=0.5$ and scale hyperparameter $B=10$ for all settings of missing data. So, we use these values of hyperparameter $A$ and $B$ for the following experiments in section~\ref{sec:LPexperiments}. In addition, we obtain that when we set $A<1$ and $B>10$, we get better results.

\subsection{Link Prediction}
\label{sec:LPexperiments}

We now compare the standard PLTF, i.e., PLTF-EM, with the proposed variational method, i.e., PLTF-VB, on a missing link prediction task. 

\subsubsection{Evaluation Metric}

In our experiments, we use Area Under the Receiver Operating Characteristic Curve (AUC) to measure the link prediction performance. Link prediction datasets are characterized by extreme imbalance, i.e., the number of links known to be present is often significantly less than the number of edges known to be absent. This issue motivates the use of AUC as a performance measure since AUC is viewed as a robust measure in the presence of imbalance \cite{StagerLT06}. The following results show the average link prediction performance of 10 independent runs in terms of AUC. 

\subsubsection{Results}

\begin{figure*}
\begin{minipage}[b]{0.33\textwidth}
\centering
  \subfigure[CP, 40\% missing]{\label{subfig:4}\includegraphics[scale=0.45]{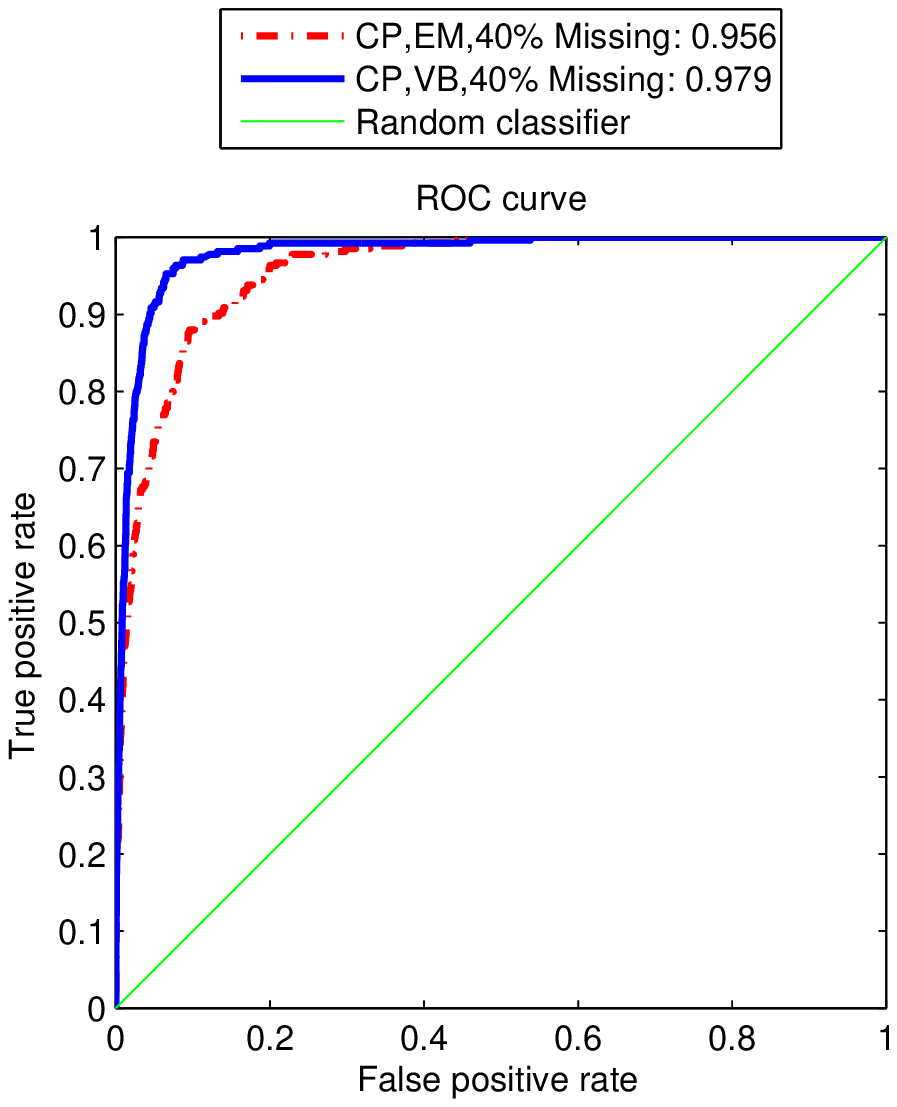}} 
\end{minipage}
\begin{minipage}[b]{0.33\textwidth}  
\centering
  \subfigure[CP, 60\% missing]{\label{subfig:5}\includegraphics[scale=0.45]{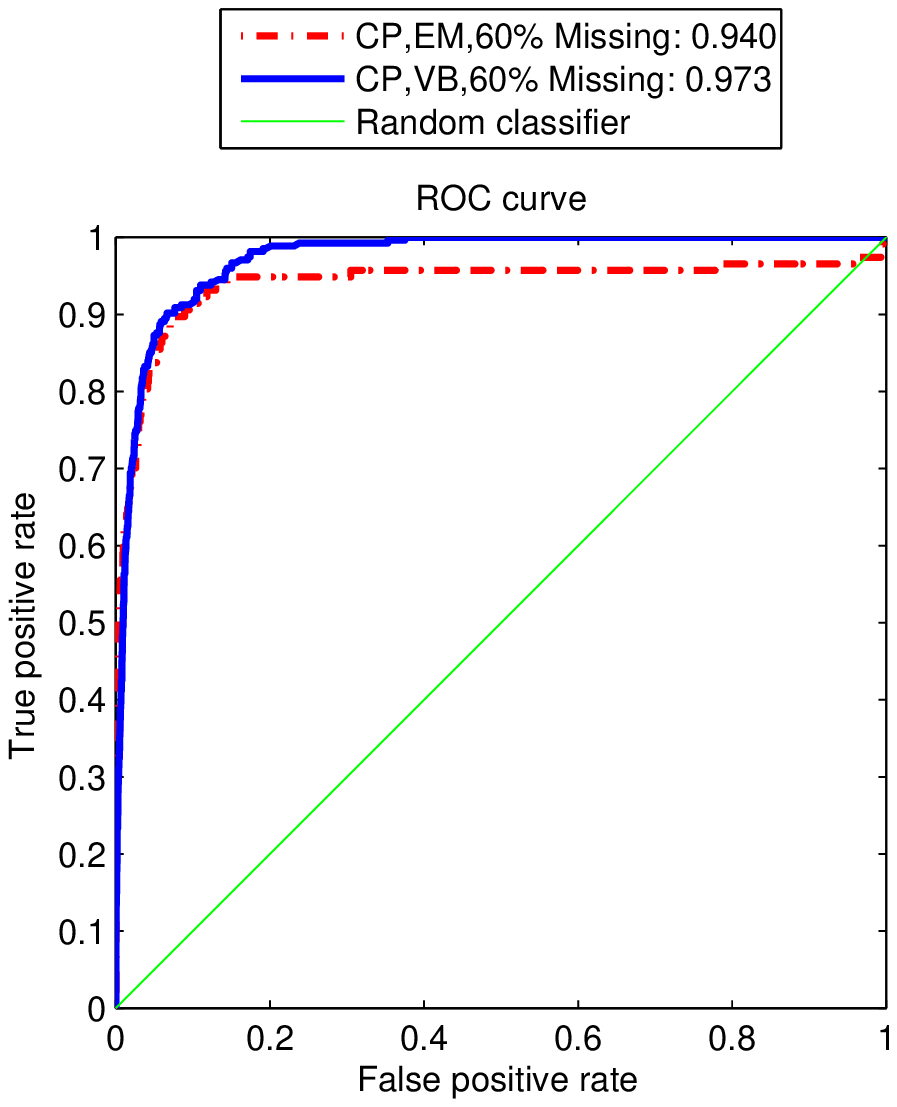}}
\end{minipage}
\begin{minipage}[b]{0.33\textwidth}
\centering
  \subfigure[CP, 80\% missing]{\label{subfig:6}\includegraphics[scale=0.45]{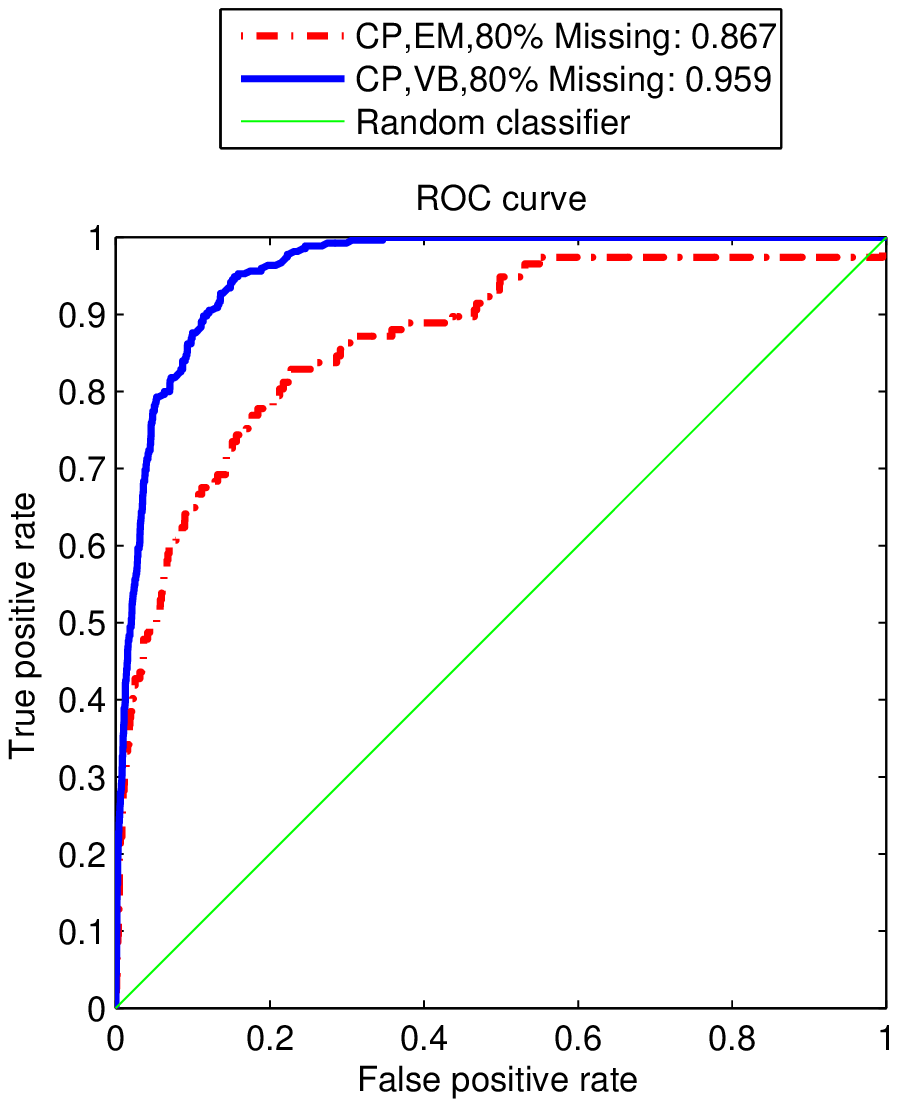}} 
\end{minipage}
\caption{Comparison of PLTF-EM and PLTF-VB methods under missing data case with CP model}
\label{fig:compVBvsEM_CP}
\end{figure*}

\begin{figure*}
\begin{minipage}[b]{0.33\textwidth}
\centering
  \subfigure[Tucker, 40\%missing]{\label{subfig:7}\includegraphics[scale=0.45]{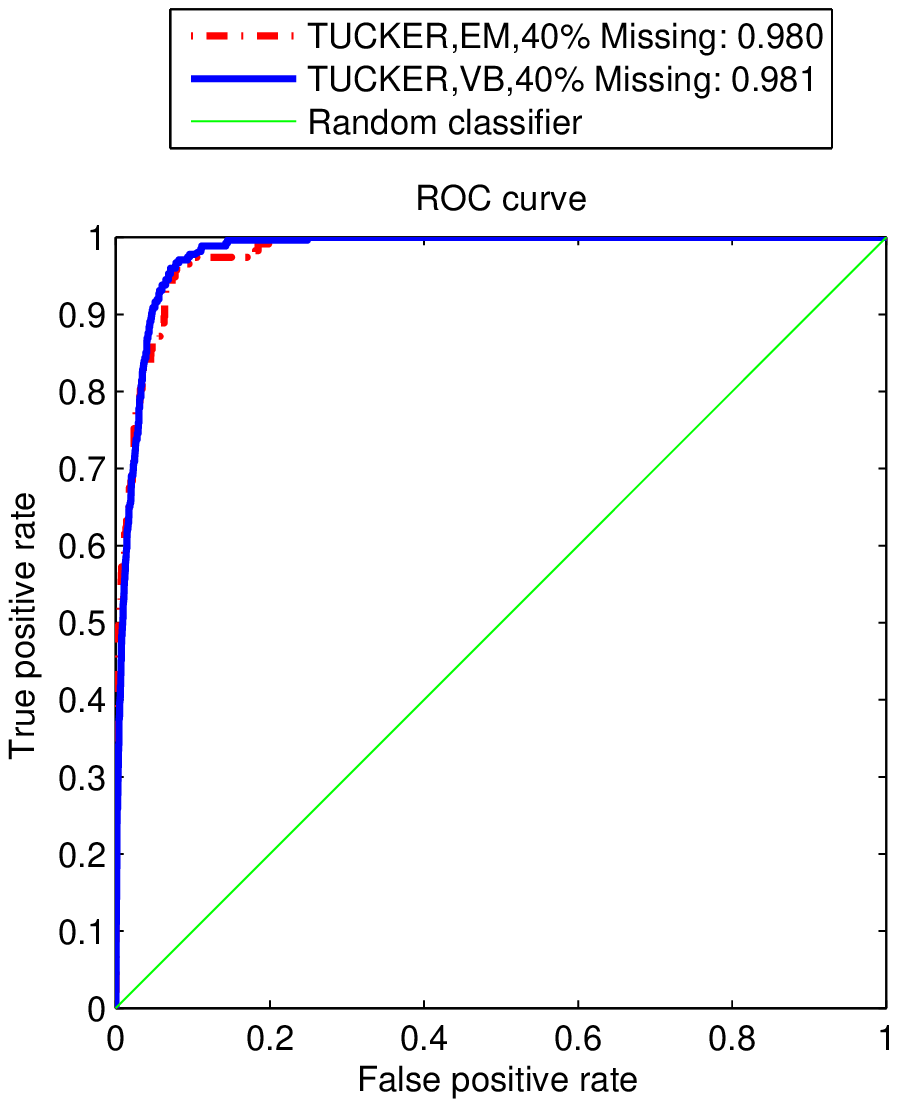}}
\end{minipage}
\begin{minipage}[b]{0.33\textwidth}  
\centering
  \subfigure[Tucker, 60\% missing]{\label{subfig:8}\includegraphics[scale=0.45]{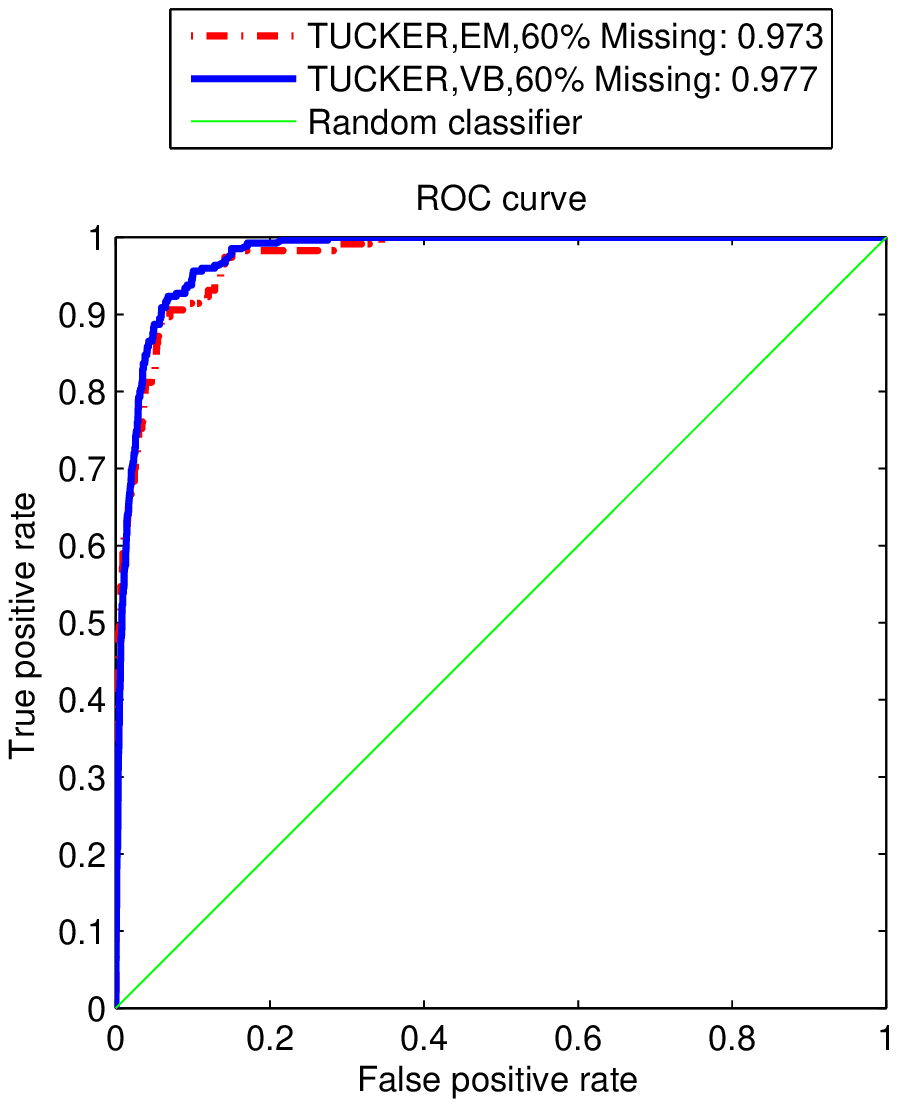}}
\end{minipage}
\begin{minipage}[b]{0.33\textwidth}
\centering
  \subfigure[Tucker, 80\% missing]{\label{subfig:9}\includegraphics[scale=0.45]{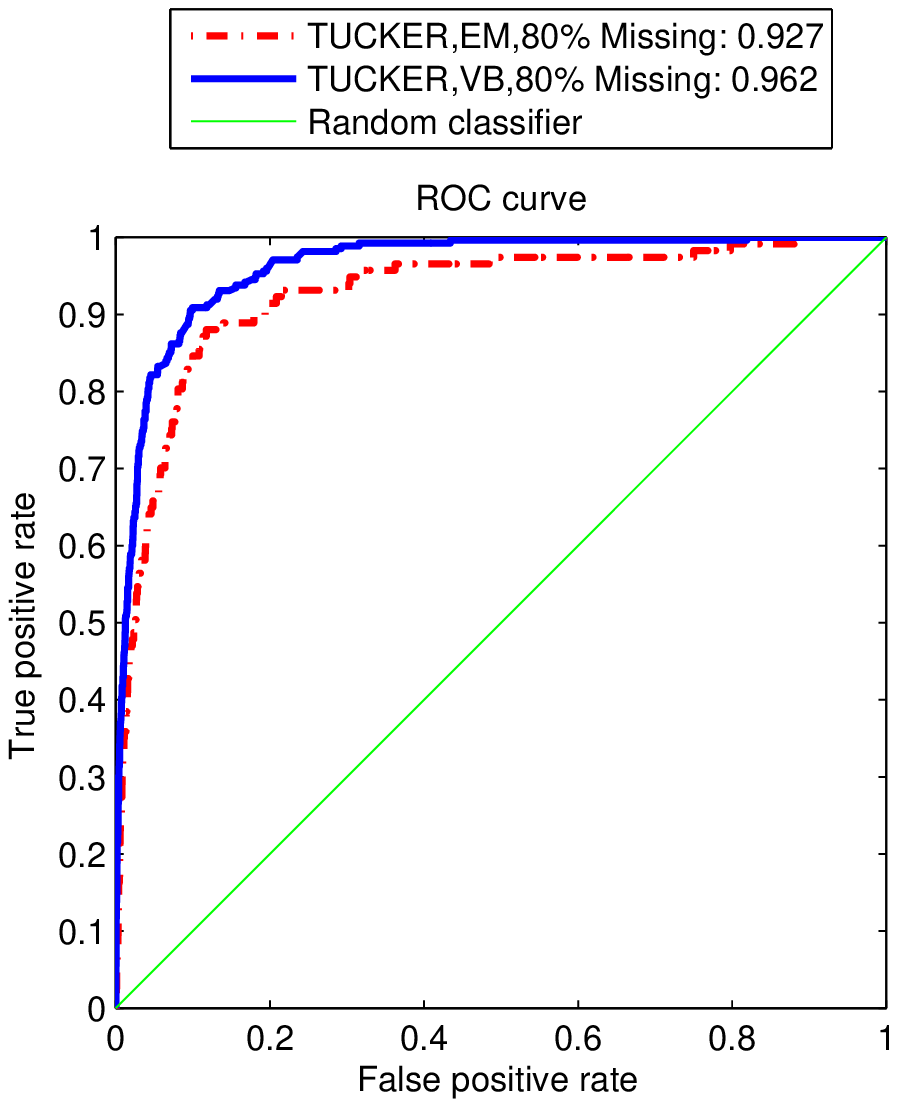}}
\end{minipage}
\caption{Comparison of PLTF-EM and PLTF-VB methods under missing data case with Tucker model}
\label{fig:compVBvsEM_Tucker}
\end{figure*}

We compare the performance of standard and variational approaches of PLTF on both CP and Tucker tensor factorization models at different amounts, i.e., $\lbrace 40, 60, 80\rbrace$, of randomly unobserved elements. In these experiments, the incomplete tensor is factorized using either a CP or a Tucker model and the extracted factor matrices are used to construct the full tensor and estimate scores for missing links. For all cases, variational approach outperforms the standard approach clearly. Figure~\ref{fig:compVBvsEM_CP} and Figure~\ref{fig:compVBvsEM_Tucker} show the comparison of PLTF-VB and PLTF-EM methods for the CP model given in Equation(\ref{eq:CP}) and the Tucker model given in Equation(\ref{eq:Tucker2}), respectively, when $\lbrace 40, 60, 80\rbrace$ of the data is missing. As we can see, the variational methods due to implicit self-regularization effect \cite{NakajimaS10}, perform better than the standard methods; in particular, when the percentage of missing data is high. Furthermore, note that the Tucker model outperforms the CP model; because Tucker model is more flexible due to the full core tensor which is helpful for us to explore the structural information embedded in the data.  

\begin{figure*}
\begin{minipage}[b]{0.33\textwidth}
\centering
  \subfigure[PLTF-VB, 40\%missing]{\label{subfig:10}\includegraphics[scale=0.45]{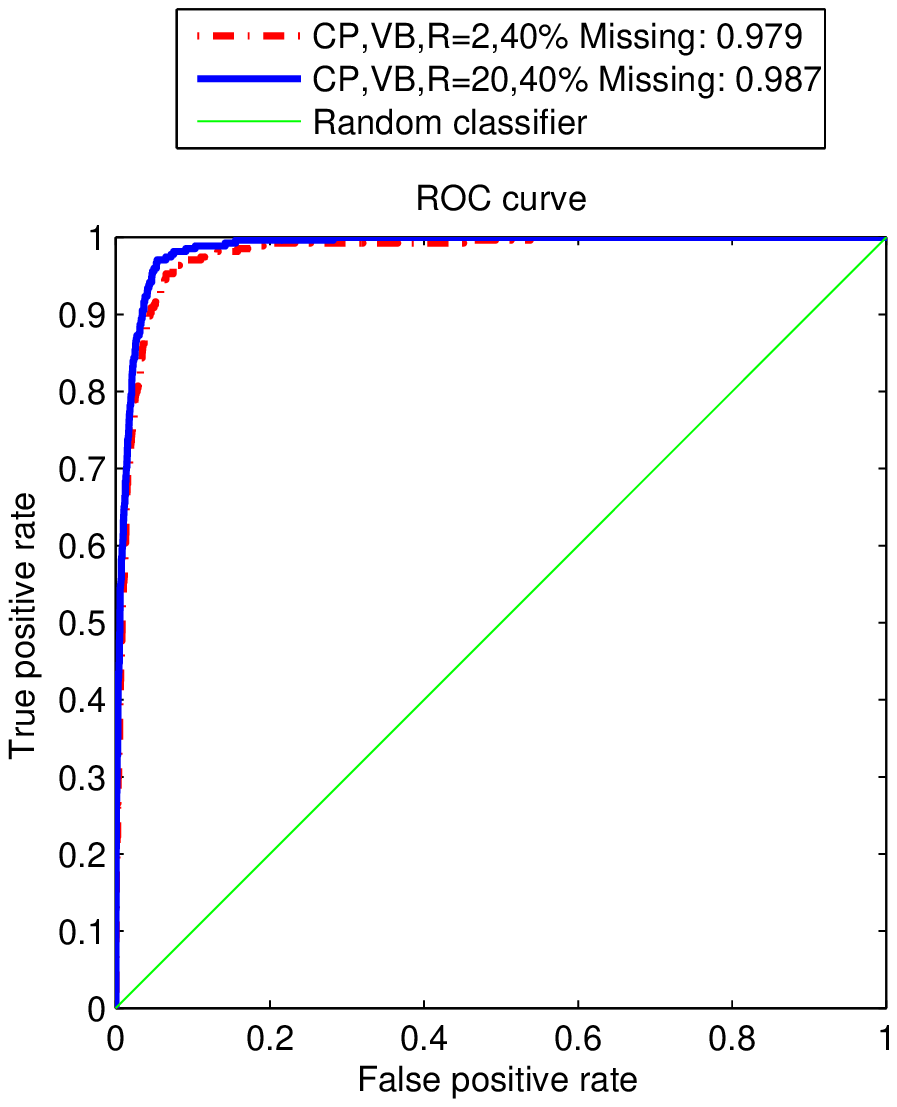}}
\end{minipage}
\begin{minipage}[b]{0.33\textwidth}  
\centering
  \subfigure[PLTF-VB, 60\%missing]{\label{subfig:11}\includegraphics[scale=0.45]{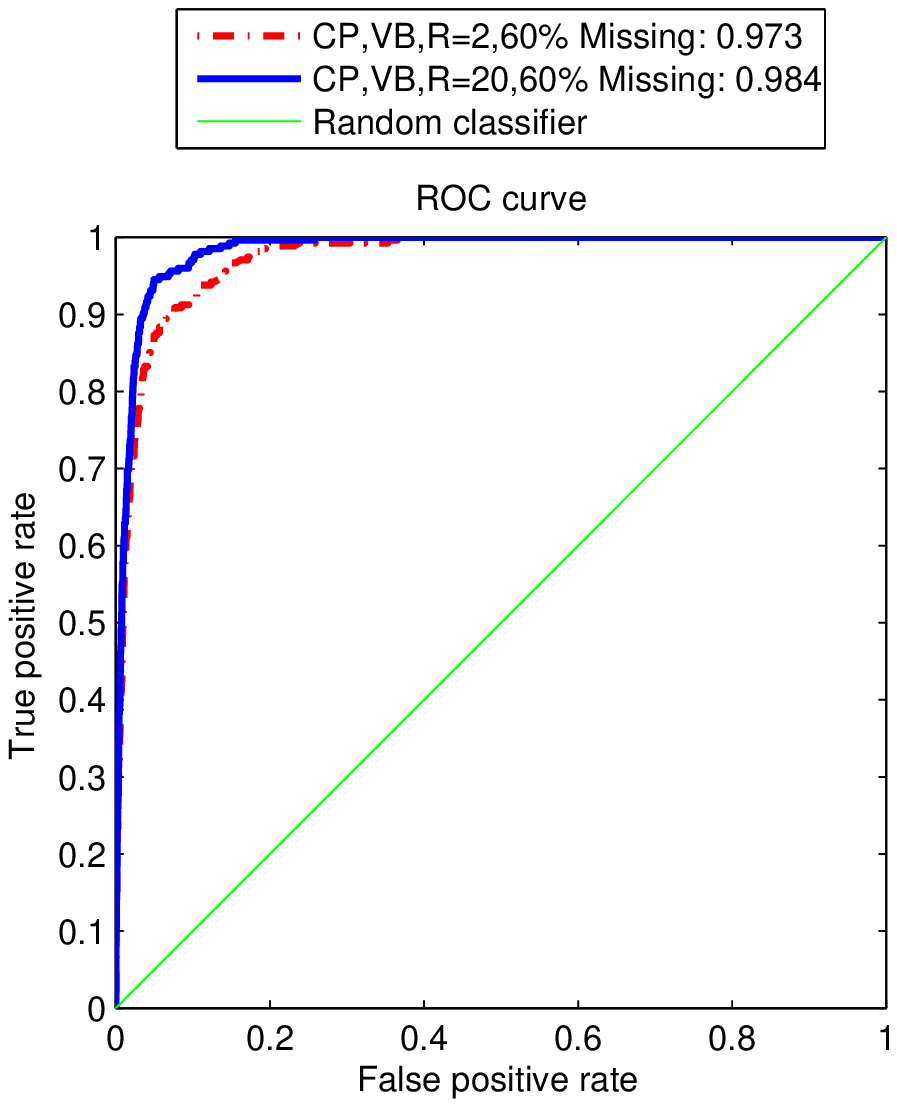}}
\end{minipage}
\begin{minipage}[b]{0.33\textwidth}
\centering
  \subfigure[PLTF-VB, 80\%missing]{\label{subfig:12}\includegraphics[scale=0.45]{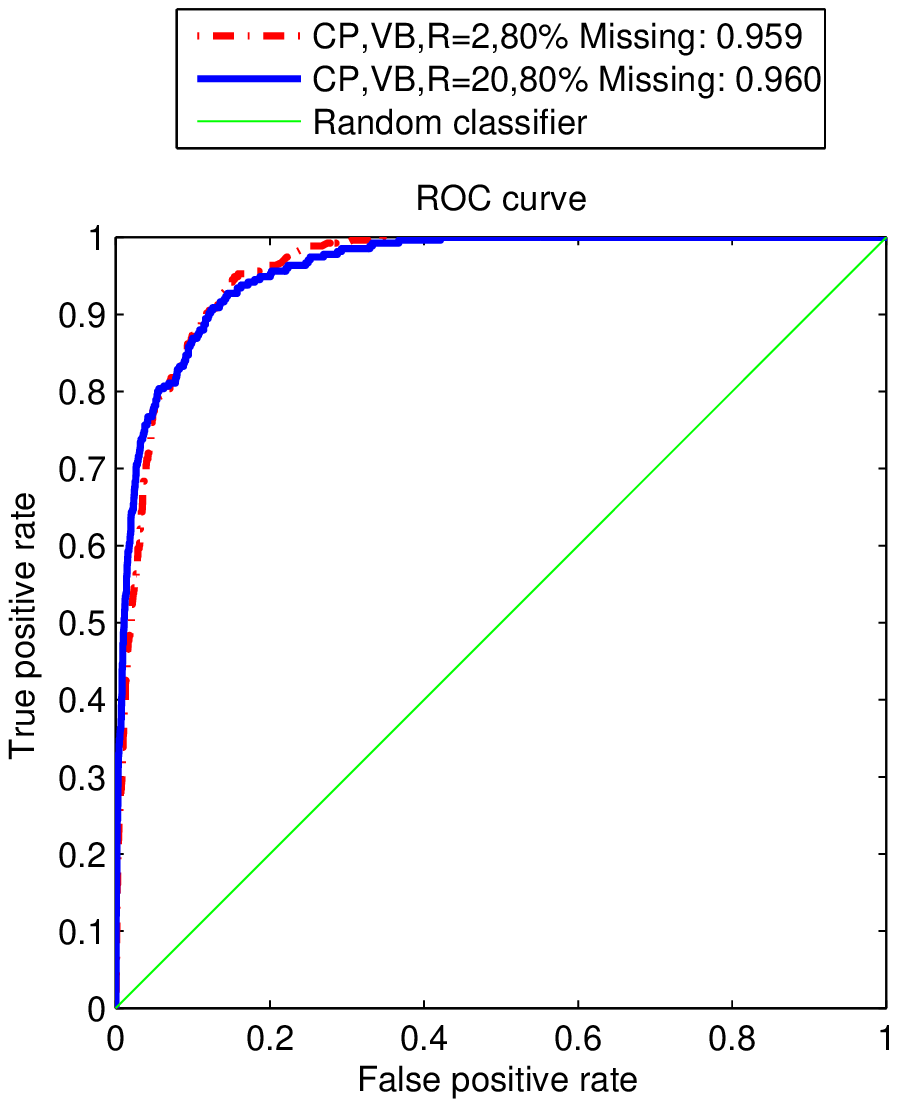}}
\end{minipage}
\caption{Effect of model order on the performance of PLTF-VB approach for CP model for different amounts of missing data.}
\label{fig:compR2vsR20_VB}
\end{figure*}

\begin{figure*}
\begin{minipage}[b]{0.33\textwidth}
\centering
  \subfigure[PLTF-EM, 40\%missing]{\label{subfig:13}\includegraphics[scale=0.45]{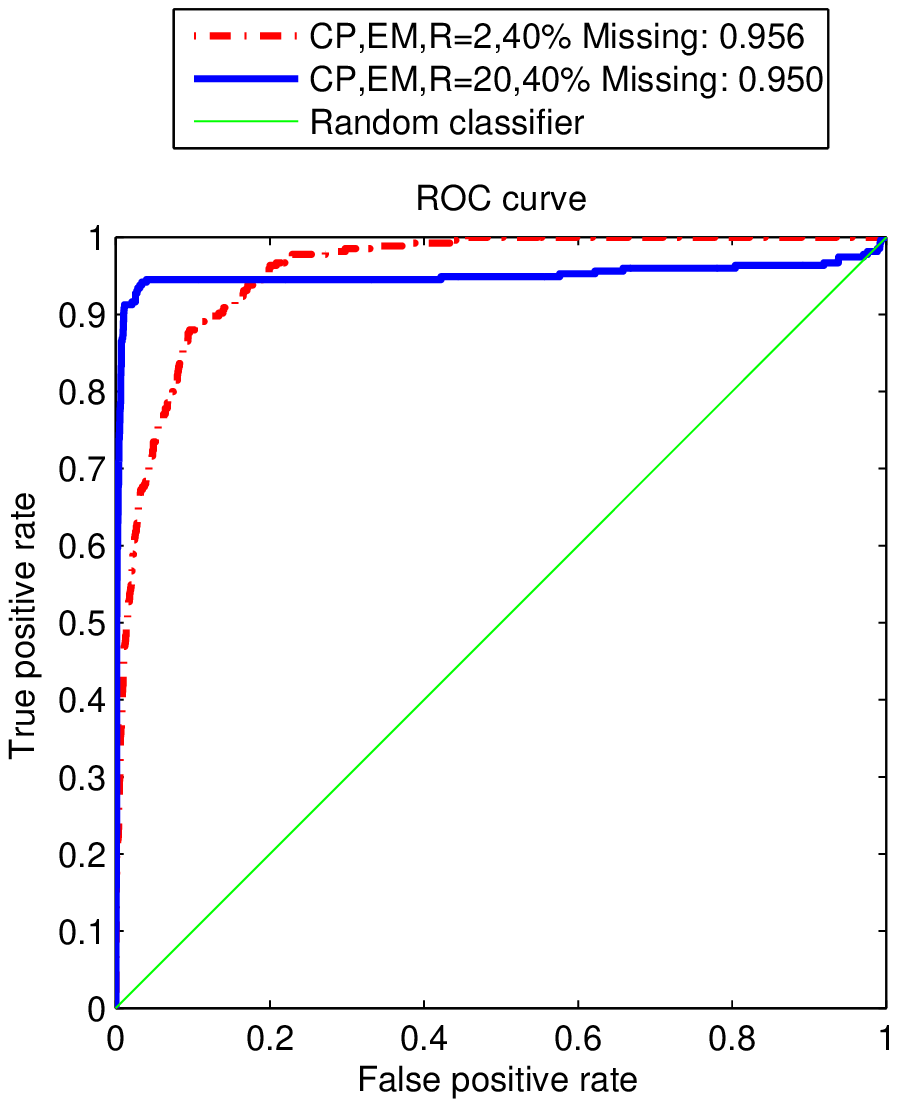}}
\end{minipage}
\begin{minipage}[b]{0.33\textwidth}  
\centering
  \subfigure[PLTF-EM, 60\%missing]{\label{subfig:14}\includegraphics[scale=0.45]{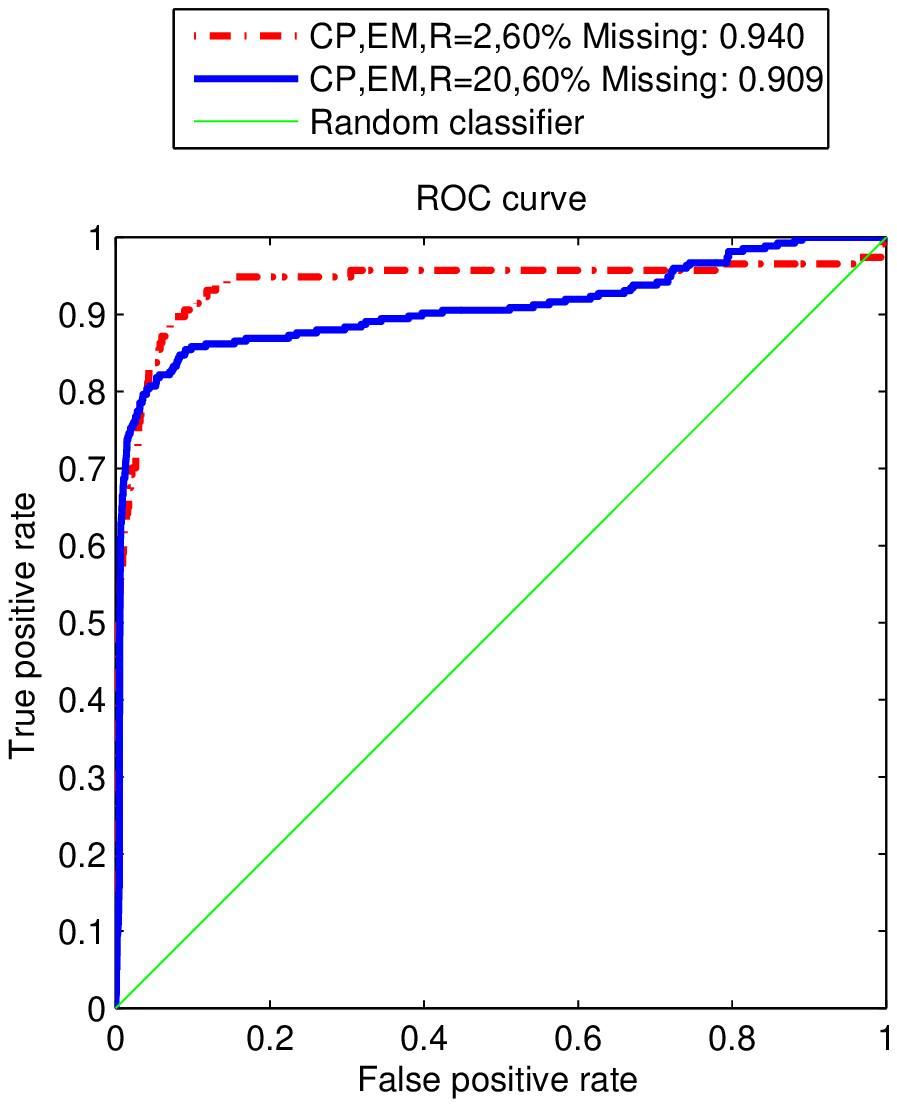}}
\end{minipage}
\begin{minipage}[b]{0.33\textwidth}
\centering
  \subfigure[PLTF-EM, 80\%missing]{\label{subfig:15}\includegraphics[scale=0.45]{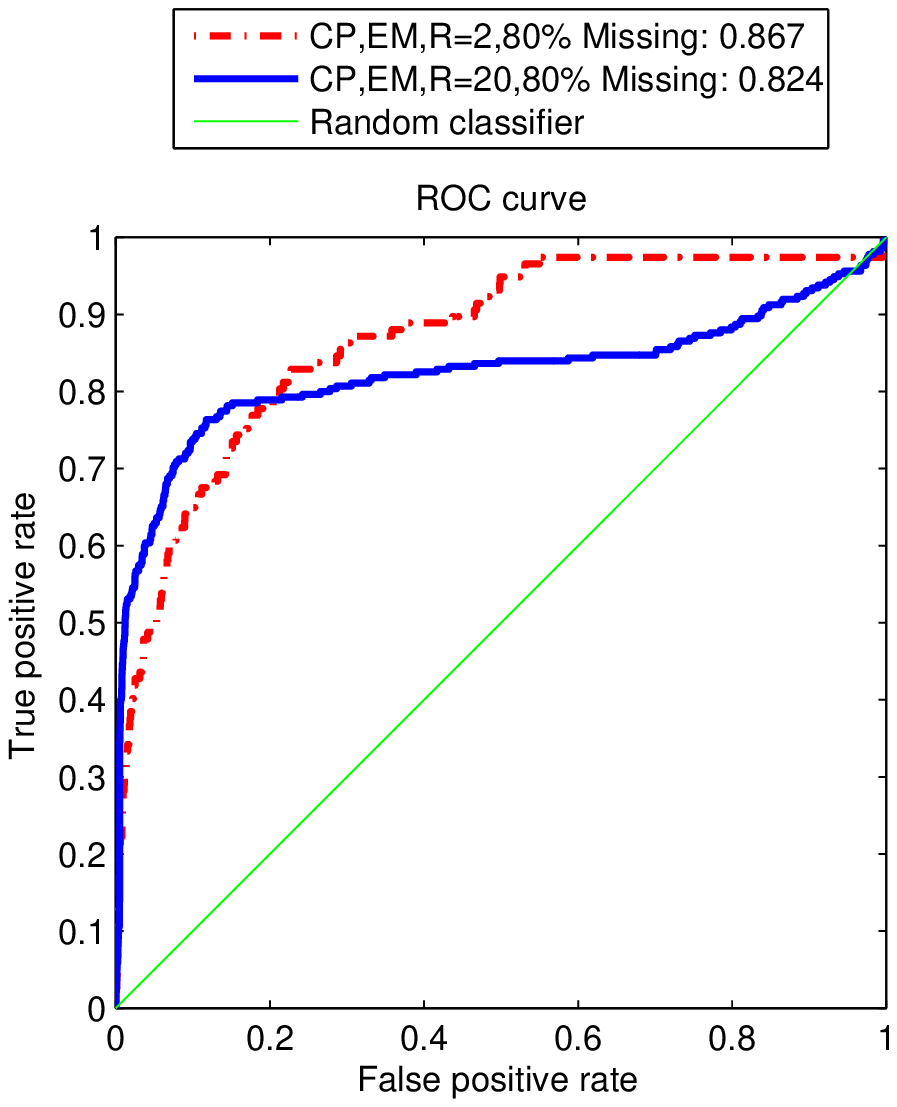}}
\end{minipage}
\caption{Effect of model order on the performance of PLTF-EM approach for CP model for different amounts of missing data.}
\label{fig:compR2vsR20_EM}
\end{figure*}

Moreover, we study the performance of PLTF-EM and PLTF-VB in terms of robustness to model order selection. As model order increases, the prediction performance of PLTF-EM drops. This is as expected since PLTF-EM is prone to overfitting and the increase in model order causes an increase in the number of free parameters that, in turn, enlarges penalty term in PLTF-EM. On the other hand, the prediction performance of the variational approach is not very sensitive to the model order and is immune to overfitting since Bayesian approach alleviates over-fitting by integrating out all model parameters \cite{cemgil09-nmf}. We compare the prediction performances of PLTF-EM and PLTF-VB methods for the CP tensor model when the component number $R$ is equal to $2$ and $20$ and for different amounts of missing data, i.e., $\lbrace 40, 60, 80\rbrace$ of the data is missing. Figure~\ref{fig:compR2vsR20_VB} and Figure~\ref{fig:compR2vsR20_EM} demonstrate that when the model order increases, the prediction performance of PLTF-VB approach stays almost same; however, the prediction performance of PLTF-EM approach declines as expected.   

\section{Related Work}
\label{sec:related}

In this section, we briefly introduce some of the related work in two categories: Bayesian inference for matrix and tensor factorizations and link prediction. 

In order to deal with the variational Bayesian matrix and tensor factorization problem, Ghahramani and Beal \cite{GhahramaniB00} provides a method that focus on deriving variational Bayesian learning in a very general form, relating it to EM, motivating parameter-hidden variable factorizations, and the use of conjugate priors. Shan et al. \cite{ProbTechreport} propose probabilistic tensor factorization algorithms, which are naturally applicable to incomplete tensors. First one is parametric probabilistic tensor factorization (PPTF), as well as a variational approximation based algorithm to learn the model and the second one is Bayesian probabilistic tensor factorization (BPTF) which maintains a distribution over all possible parameters by putting a prior on top, instead of picking one best set of model parameters. Cemgil \cite{cemgil09-nmf} describes a non-negative matrix factorization (NMF) in a statistical framework, with a hierarchical generative model consisting of an observation and a prior component. Starting from this view, he develops full Bayesian inference via variational Bayes or Monte Carlo.

Nakajima et al. \cite{NakajimaST10} propose a global optimal solution to variational Bayesian matrix factorization (VBMF) that can be computed analytically by solving a quartic equation and it is highly advantageous over a popular VBMF algorithm based on iterated conditional modes (ICM), since it can only find a local optimal solution after iterations. Yoo and Choi \cite{YooC11} present a hierarchical Bayesian model for matrix co-factorization in which they derive a variational inference algorithm to approximately compute posterior distributions over factor matrices. 

For Bayesian model selection, Sato \cite{Sato01} derives an online version of the variational Bayes algorithm and proves its convergence by showing that it is a stochastic approximation for finding the maximum of the free energy. By combining sequential model selection procedures, the online variational Bayes algorithm provides a fully online learning method with a model selection mechanism. 

We next turn to link prediction studies. Most often, an incomplete set of links is observed and the goal is to predict unobserved links (also referred to as the \emph{missing link prediction} problem), or there is a temporal aspect: snapshots of the set of links up to time $t$ are given and the goal is to predict the links at time $t+1$ (\emph{temporal link prediction} problem). Matrix and tensor factorization-based methods have recently been studied for temporal link prediction \cite{AcDuKo10b}; however, in this paper, we have considered the use of tensor factorizations for the missing link prediction problem. Applications of missing link prediction include predicting  links in social networks \cite{Aaron2008}; predicting the participation of users in events such as email communications and co-authorship \cite{GeDi05} and predicting the preferences of users in online retailing \cite{KoBeVo09}. Matrix factorization and tensor factorization-based approaches have proved useful in terms of missing link prediction because missing link prediction is closely related to matrix and tensor completion studies, which have shown that by using a low-rank structure of a data set, it is possible to recover missing entries accurately for matrices \cite{CaPl09} and higher-order tensors \cite{AcDuKoMo10,GaReYa11}.

\section{Conclusions}
\label{sec:conc}

In this paper, we have investigated variational inference for PLTF framework with KL cost from a full Bayesian perspective that also handles the missing data naturally. In addition, we develop a practical way without incurring much additional computational cost to PLTF-EM approach for computing the approximation distribution and full conditionals; then, we estimate the model order in terms of marginal likelihood. By maximizing the bound on marginal likelihood, we have a method where all the hyperparameters can be estimated from data. Our experiments suggest that the variational bound seems to be reasonable approximation to the marginal likelihood and can guide model selection for PLTF. 

As a future direction and next step of this work, we aim to extend our variational method in order to be able to make inference on tensor factorization models where multiple observed tensors $(X_1, ..., X_K)$ can share a set of factors \cite{yilmazGTF}. Factorization of multiple observed tensors simultaneously, alleviates the overfitting better than the standard variational Bayesian matrix factorization and leads to the improved performance \cite{YooC11}.

\bibliographystyle{splncs}
\bibliography{paper}

\end{document}